# Ionization Energies, Electron Affinities, Bandgaps, Exciton Binding Energies, and Polarization Energies of Orientation-Controlled Picene, [6]-Phenacene, and [7]-Phenacene Thin Films

Rintaro Makino[1], Mihiro Kubo[1], Keiichirou Yonezawa[2], Hiroyuki Yoshida[1,3*], Satoshi Kera[1,4,5]

1. Graduate School of Engineering, Chiba University, 1-33 Yayoi-cho, Inage-ku, Chiba 263-8522, Japan
2. Industry-Academia Innovation Center, Kyushu Institute of Technology, Kitakyushu-shi, Fukuoka, 804-8550, Japan
3. Molecular Chirality Research Center, 1-33 Yayoi-cho, Inage-ku, Chiba University, Chiba 263-8522, Japan
4. Institute for Molecular Science, National Institutes of Natural Sciences, Okazaki 444-8585, Japan
5. The Graduate University for Advanced Studies (SOKENDAI), NishigoNaka 38, Myodaiji, Okazaki 444-8585, Japan

*corresponding author: hyoshida@chiba-u.jp

**Abstract**

Phenacenes, in which benzene rings are fused in a zigzag manner, are chemically robust π-conjugated hydrocarbons of interest for organic electronic applications and superconductivity. However, quantitative electronic parameters of phenacene thin films remain limited compared with those of acenes. Here, we determine the ionization energies (IEs), electron affinities (EAs), band gaps, exciton binding energies, and polarization energies of orientation-controlled thin films of picene, [6]-phenacene, and [7]-phenacene using ultraviolet photoelectron spectroscopy and low-energy inverse photoelectron spectroscopy. The energy parameters depend only weakly on molecular size. In contrast, both IE and EA exhibit large orientation-dependent shifts of approximately 1 eV: lying films show IEs of 6.41–6.43 eV and EAs of 2.16–2.42 eV, whereas standing films show IEs of 5.44–5.67 eV and EAs of 1.36–1.53 eV. Nevertheless, the band gaps of approximately 4 eV and exciton binding energies of approximately 1 eV remain nearly independent of molecular size and orientation. Analysis of polarization energies shows that the induction term is nearly orientation independent, whereas the electrostatic term strongly depends on orientation and originates primarily from molecular quadrupole moments. These results identify molecular orientation as a key factor governing frontier energy levels in phenacene thin films.

**Keywords:** phenacene; picene; molecular orientation; ultraviolet photoelectron spectroscopy (UPS); low-energy inverse photoelectron spectroscopy (LEIPS); ionization energy; electron affinity; bandgap; polarization energy;

## 1. Introduction

Organic semiconductors offer a unique platform for electronic devices that combine molecular designability with mechanical flexibility, light weight, and low-temperature processability. In organic field-effect transistors (OFETs) and related organic electronic devices, device operation is governed not simply by the molecular orbital energies of isolated molecules, but by the frontier transport levels formed in the solid state and at interfaces. The hole and electron transport levels are represented by the ionization energy (IE) and electron affinity (EA), respectively, which correspond to the edges of the highest occupied molecular orbital (HOMO)- and lowest unoccupied molecular orbital (LUMO)-derived transport states referenced to the vacuum level. Reliable values of IE and EA, as well as bandgap (transport gap), are therefore essential for evaluating fundamental semiconductor properties, hole and electron injection barriers, and contact resistance. For optoelectronic applications, the exciton binding energy also plays an important role.

Polycyclic aromatic hydrocarbons (PAHs), which are composed of fused benzene rings, are one of the most fundamental classes of organic semiconductors. PAHs can be regarded as finite molecular substructures of graphene. This viewpoint also has historical roots in early pioneering studies of organic semiconductors in the 1940s and 1950s by Akamatsu and Inokuchi, in which conducting graphite motivated investigations of PAH molecular crystals [1, 2]. PAHs can be classified into several families (see Figure 1), including acenes, in which benzene rings are linearly annulated; phenacenes, in which benzene rings are fused in a zigzag manner; and helicenes, which are non-planar, helically twisted polycyclic benzenoid molecules [3]. Owing to the structural similarity between PAHs and graphene nanoribbons in terms of their local benzenoid building blocks, their electronic structures are strongly affected by molecular topology and edge geometry. For graphene nanoribbons, for example, the bandgap depends sensitively on whether the edge structure is armchair or zigzag [4]. Linearly fused acenes can be viewed as molecular fragments related to armchair-type graphene nanoribbons, whereas phenacenes, which consist of benzene rings fused in a zigzag manner, represent complementary zigzag benzenoid frameworks.

Among these PAHs, acenes have been studied most extensively. The HOMO–LUMO gap of acenes decreases rapidly as the number of fused benzene rings increases. Consequently, higher acenes such as hexacene and longer homologues are unstable in air, although they can be generated from air-stable precursors and their electronic structures have been investigated [5]. Pentacene, which consists of five linearly fused benzene rings, has therefore been studied extensively as a benchmark organic semiconductor. Pentacene field-effect transistors have been investigated in both thin-film and single-crystal devices [6-11]. From a fundamental perspective, pentacene has also been widely studied in terms of its valence-band electronic structure [12-21], conduction-band electronic structure [22], and solid-state structure, including polymorphs in both single crystals [23-28] and thin films [29-32]. With respect to frontier energy levels, pentacene has also served as a benchmark material for investigating orientation-dependent IE and EA [33-35].

In contrast to acenes, phenacenes exhibit distinct electronic properties: they retain relatively large HOMO–LUMO gaps even as the π-system is extended [36-38]. Their deep frontier energy levels make them less susceptible to oxidation, contributing to their high chemical stability [39]. Regarding OFET applications, high-performance thin-film and single-crystal FETs based on picene ([5]-phenacene) [40, 41], [6]-phenacene [42], and [7]-phenacene [43] have been reported. In addition, electron-doped phenacene solids have been reported to exhibit superconductivity [44, 45].

Despite these remarkable semiconductor properties, basic electronic-level data for phenacene films remain insufficient. The valence-band structure of picene single crystals has previously been reported [46]. With respect to frontier energy levels, the IE measured by ultraviolet photoelectron spectroscopy and the EA measured by conventional inverse photoemission spectroscopy were reported for picene and [7]-phenacene films [37]. However, the bandgaps derived from these data are 3.2 eV for picene and 3.1 eV for [7]-phenacene, which are almost identical to their respective optical gaps of 3.3 and 3.1 eV. These values are unlikely, because the bandgap is usually larger than the optical gap by 0.3–1.0 eV owing to the exciton binding energy [47-50]. This discrepancy may be partly attributable to the use of conventional IPES for EA determination, which is intrinsically less reliable for organic materials because incident electrons can damage the sample and because the energy resolution is not sufficiently high to determine the EA accurately. Furthermore, the reported energy levels for the films did not take molecular orientation into consideration. This distinction is crucial because the IE and EA of organic molecular films can vary substantially with molecular orientation [35, 51-53].

In this study, we address these issues by preparing orientation-controlled thin films of picene, [6]-phenacene, and [7]-phenacene on highly oriented pyrolytic graphite (HOPG) and naturally oxidized silicon ($SiO_2$) substrates. A combination of UPS and metastable atom electron spectroscopy (MAES) [54] is used to examine the molecular orientations. Gas-phase UPS provides the isolated-molecule IE reference, while thin-film UPS determines the solid-state IE. Low-energy inverse photoelectron spectroscopy (LEIPS) [55-57] is used to determine the EA. The use of LEIPS is central to this work because it provides less damaging measurements and more reliable EA values than conventional IPES, allowing direct comparison of orientation-dependent energy parameters derived from IE and EA. By systematically examining picene, [6]-phenacene, and [7]-phenacene, we clarify how the frontier transport levels, namely IE and EA, as well as the bandgaps, exciton binding energies, and polarization energies, evolve with π-conjugation length and molecular orientation. The results show that both IE and EA are strongly orientation dependent, whereas the bandgap is almost unchanged in phenacene thin films. By evaluating the polarization energies, we elucidate the origin of the orientation dependence of these energy parameters.

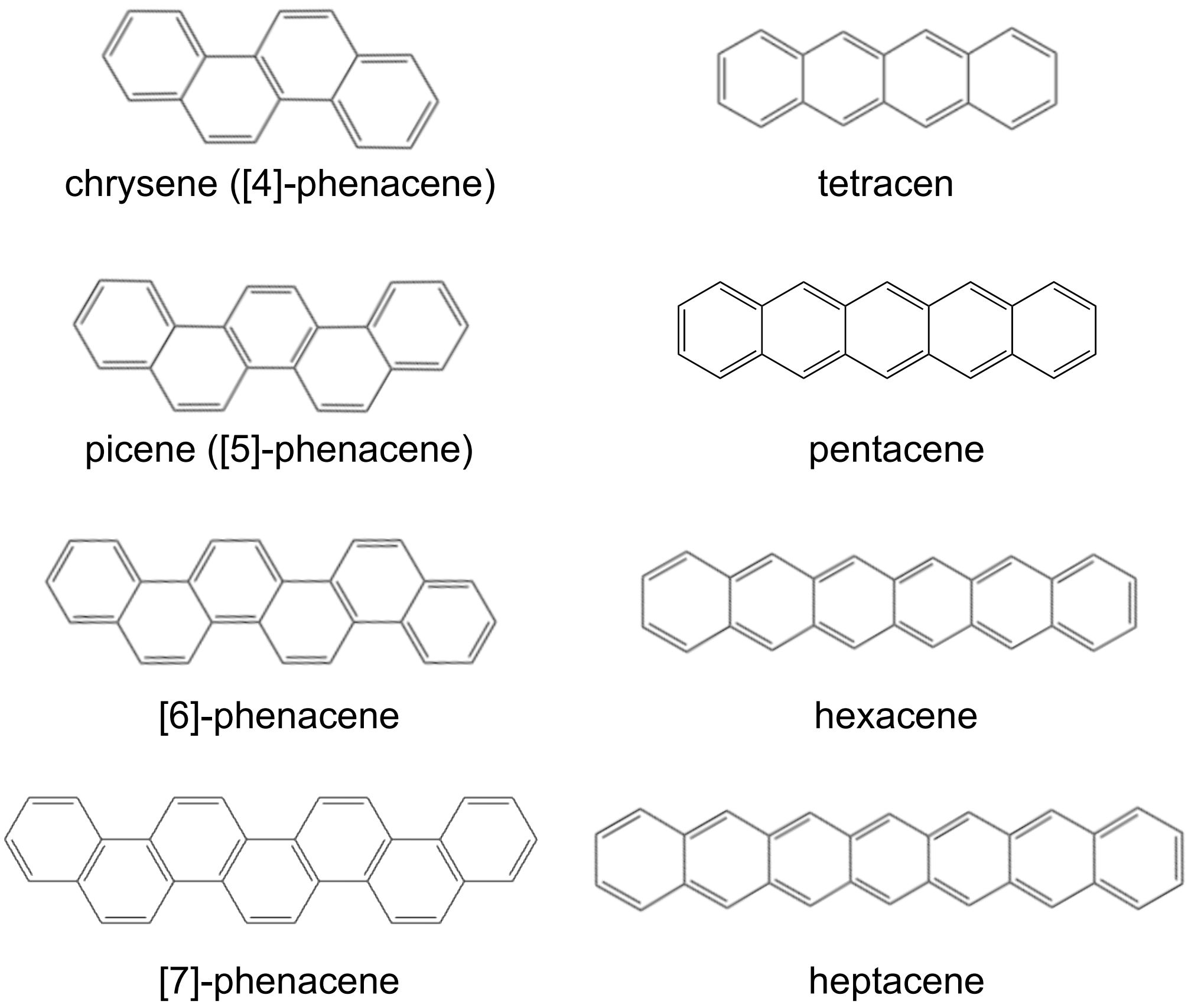

**Figure 1** Molecular structures of phenacenes and the corresponding acenes. Molecular structures of the phenacenes examined in this study are shown together with acene molecules containing the same number of fused benzene rings.

## 2. Results and Discussion

### 2.1 Molecular orientations of phenacenes on $SiO_2$ and HOPG

The thin-film structures of picene on $SiO_2$ and HOPG have been investigated in detail using X-ray reflectivity/diffraction and electron spectroscopies [58]. For [6]-phenacene and [7]-phenacene on $SiO_2$, thin-film structures have also been reported based on X-ray diffraction measurements [59]. In both cases, the molecules adopt a standing (or edge-on) orientation on $SiO_2$. In contrast, picene adopts a lying (or face-on) orientation on HOPG. Such substrate-dependent molecular orientations are commonly observed for organic semiconductors, such as phthalocyanines [52], diindenoperylene (DIP) [60], hexa-peri-hexabenzocoronene (HBC) [61], pentacene [35], pentacenequinone [53], and perylene diimides [62].

To determine the molecular orientation of the phenacene thin films, we combined ultraviolet photoelectron spectroscopy (UPS) with metastable atom electron spectroscopy (MAES). While UPS probes the occupied electronic states of the film, MAES is highly surface sensitive and preferentially detects molecular orbitals exposed at the outermost surface. Figure 2(a) shows the calculated density-of-states spectrum and molecular orbital energies of isolated picene at the B3LYP/cc-pVDZ level, together with the gas-phase UPS spectrum and the UPS and MAES spectra of picene films deposited on HOPG and $SiO_2$. For the orientation analysis, we focus on representative molecular orbitals associated with selected spectral bands: band A (HOMO and HOMO-1), band B (HOMO−7 and HOMO-8), band C (HOMO−9, HOMO-10 and HOMO-11), and band D (HOMO−16, HOMO-17, HOMO-18 and HOMO-20).). As illustrated in Fig. 2(b), the π-type orbitals, represented by the HOMO and HOMO−7, extend predominantly out of the molecular plane, whereas the σ-type orbitals, represented by HOMO−9 and HOMO−16, are localized mainly along the molecular framework and terminal regions. Consequently, as schematically illustrated in Figure 2(c), enhancement of the π-derived bands in MAES indicates a lying-flat (face-on) orientation, whereas enhancement of the σ-derived bands indicates a standing (edge-on) orientation. Consistent with this picture, the MAES spectrum of picene on HOPG shows enhanced π-derived bands A and E, indicating that the molecular plane is nearly parallel to the substrate. In contrast, the σ-derived bands I and J are enhanced for picene on $SiO_2$, demonstrating that the molecules adopt a standing orientation with their long molecular axes approximately normal to the substrate surface.

The same analysis was applied to [6]-phenacene and [7]-phenacene, whose UPS and MAES spectra are shown in Figures 2(d) and 2(e). The spectral features were assigned by comparison with DFT-calculated molecular orbital energies at the B3LYP/6-31+G(d) level. For [6]-phenacene, bands A (HOMO, HOMO−1, and HOMO−2) and B (HOMO−7 and HOMO−9) are predominantly π-derived, whereas bands C (HOMO−14) and D (HOMO−19 and HOMO−20) are predominantly σ-derived. For [7]-phenacene, bands A (HOMO−3 and HOMO−4) and B (HOMO−6 and HOMO−7) are π-derived, while bands C (HOMO−12, HOMO−13, and HOMO−14) and D (HOMO−22, HOMO−23, and HOMO−25) are σ-derived. In both [6]- and [7]-phenacene films on $SiO_2$, the σ-derived features are enhanced in the MAES spectra, indicating a standing (edge-on) molecular orientation, consistent with previous X-ray diffraction studies.[59] In contrast, the π-derived features are preferentially enhanced for films grown on HOPG, confirming a lying-flat (face-on) orientation. In particularly, the σ-derived bands C and D are clearly resolved only in MAES spectra of both materials on SiO2, providing further evidence for the standing orientation. Thus, all three phenacenes exhibit the same substrate-dependent orientation, standing on $SiO_2$ and lying flat on HOPG.

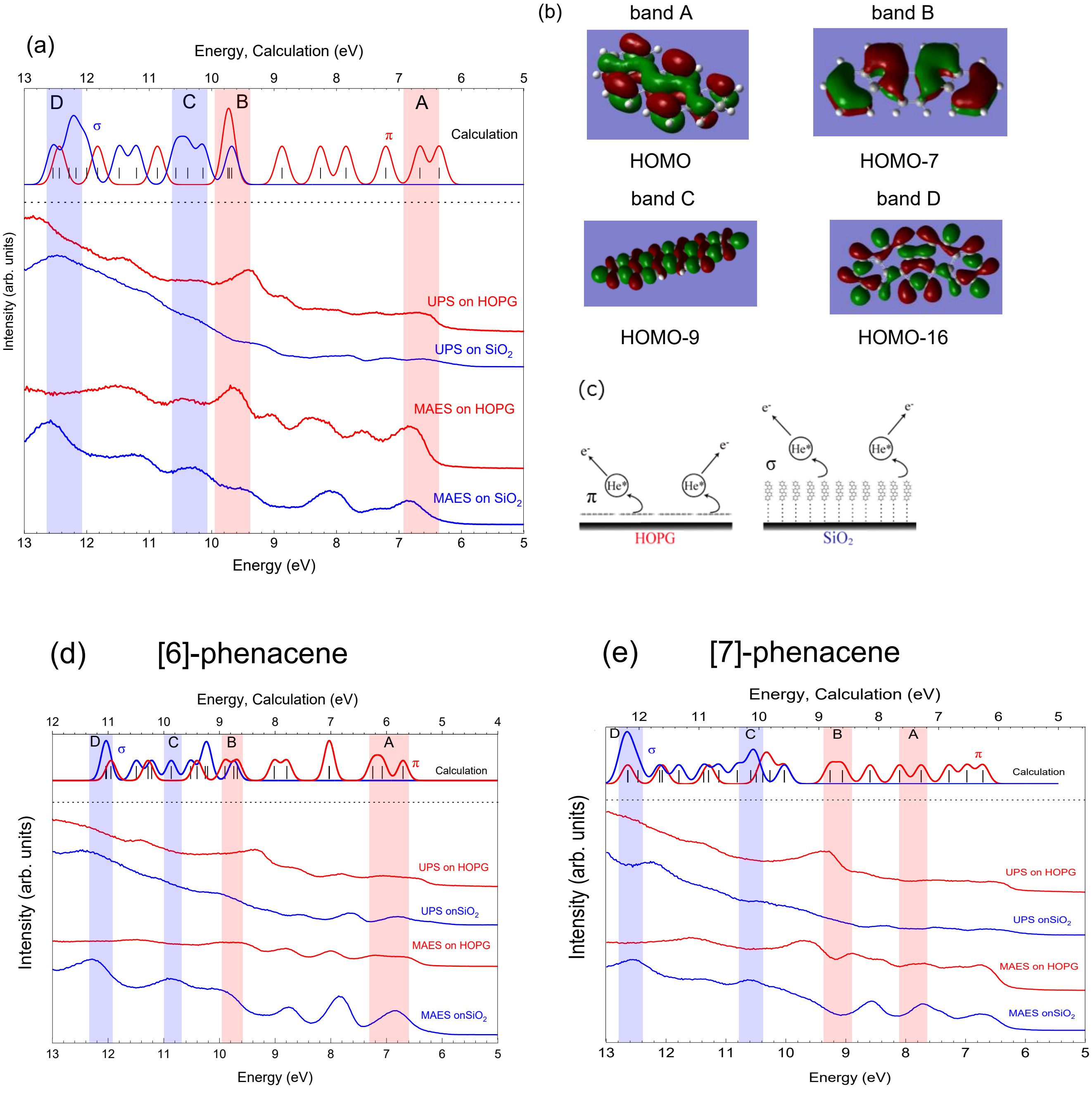

**Figure 2** Molecular orientation analysis by UPS and MAES. (a) Calculated molecular orbital energies, gas-phase UPS, and UPS/MAES spectra of picene films on HOPG and $SiO_2$. (b) Representative orbitals assigned to the spectral bands: HOMO, HOMO−7, HOMO−9, and HOMO−16 of picene. (c) Schematic of the MAES/UPS-based molecular orientation analysis. Enhanced π-orbital bands indicate a lying orientation on HOPG, whereas enhanced σ-orbital bands indicate a standing orientation on $SiO_2$. (d) Calculated molecular orbital energies, and UPS/MAES spectra of films on HOPG and $SiO_2$ for (d) [6]-phenacene and (e) [7]-phenacene.

**2.2 UPS and LEIPS of thin films: ionization energy, electron affinity, bandgap, and exciton binding energy**

Figure 3 shows the combined UPS and LEIPS spectra with the energy referenced to the vacuum level. IE and EA values were determined from the onsets of the HOMO-derived UPS feature and the LUMO-derived LEIPS feature, respectively. For UPS, the observed spectral line shape and IE are consistent with those reported previously for picene [58] and [6]-phenacene [63], supporting the reliability of the present experiment. The thin-film IE and EA values are almost independent of molecular size from picene to [7]-phenacene, but reveal strong orientation effects as summarized in Table 1. For standing films on $SiO_2$, IE values are 5.67, 5.49, and 5.44 eV for picene, [6]-phenacene, and [7]-phenacene, respectively. For lying films on HOPG, the corresponding values are 6.43, 6.42, and 6.41 eV. Thus, changing the orientation from standing to lying increases IE by 0.76 eV for picene, 0.93 eV for [6]-phenacene, and 0.97 eV for [7]-phenacene. The EA values measured by LEIPS show the same direction of orientation dependence. The standing films on $SiO_2$ have EA values of 1.52, 1.36, and 1.53 eV, whereas lying films on HOPG have EA values of 2.16, 2.30, and 2.42 eV for picene, [6]-phenacene, and [7]-phenacene, respectively. The EA increase from the standing to the lying orientation is 0.64, 0.94, and 0.89 eV.

From the obtained values, the bandgap was calculated as IE - EA. The bandgap is much less orientation-sensitive than either frontier level alone. The standing films have bandgaps of 4.15 eV for picene, 4.13 eV for [6]-phenacene, and 3.91 eV for [7]-phenacene, whereas the lying films have corresponding values of 4.27, 4.12, and 3.99 eV. The gap differences between lying and standing films are therefore only 0.12, -0.01, and 0.08 eV. Thus, molecular orientation can shift both charge-transport levels by nearly 1 eV while leaving the bandgap almost unchanged. This behavior is directly relevant to device energetics because injection barriers can be changed by molecular orientation without substantially changing the intrinsic charge-separation energy.

The difference between the bandgap and the optical gap is interpreted as the exciton binding energy. The exciton binding energies estimated in this study are approximately 1 eV (Table 1) and are again almost the same among the three compounds. It is interesting to note that the exciton binding energy corresponds to about one-third of the optical gaps (~3 eV) and one-quarter of the bandgaps (~4 eV). This finding is consistent with our previous results on the relationship between exciton binding energy and bandgap in organic molecular solids [50].

Previous electronic-structure studies of phenacenes used PES/IPES to estimate the occupied and unoccupied levels of selected thin films, namely picene and [7]-phenacene on Au/ITO [36, 37]. In those studies, the HOMO and LUMO levels relative to the Fermi level were reported to be −1.83 and +1.43 eV, respectively, for picene, and −1.68 and +1.44 eV, respectively, for [7]-phenacene. The resulting gaps were 3.26 eV for picene and 3.12 eV for [7]-phenacene. These values are close to the optical gaps and are about 1 eV smaller than the bandgaps obtained in the present work. If these PES/IPES gaps are interpreted as bandgaps, the corresponding exciton binding energies would be nearly zero, which is unlikely for molecular solids of this size [50]. One possible reason for this discrepancy is that the previous studies employed conventional IPES, in which sample damage may not be negligible and the energy resolution is limited. Uncertainty in the LUMO-onset position in conventional IPES can directly affect the estimated EA and bandgap. Another possible reason is that the previous values were determined with respect to the Fermi level. For organic semiconductors, however, the band gap is more appropriately determined from the HOMO and LUMO onsets referenced to the vacuum level, that is, from the difference between the IE and EA [47, 50, 64]. The present LEIPS measurements, together with vacuum-level referencing, address these issues and enable reliable evaluation of the band gaps of orientation-controlled films.

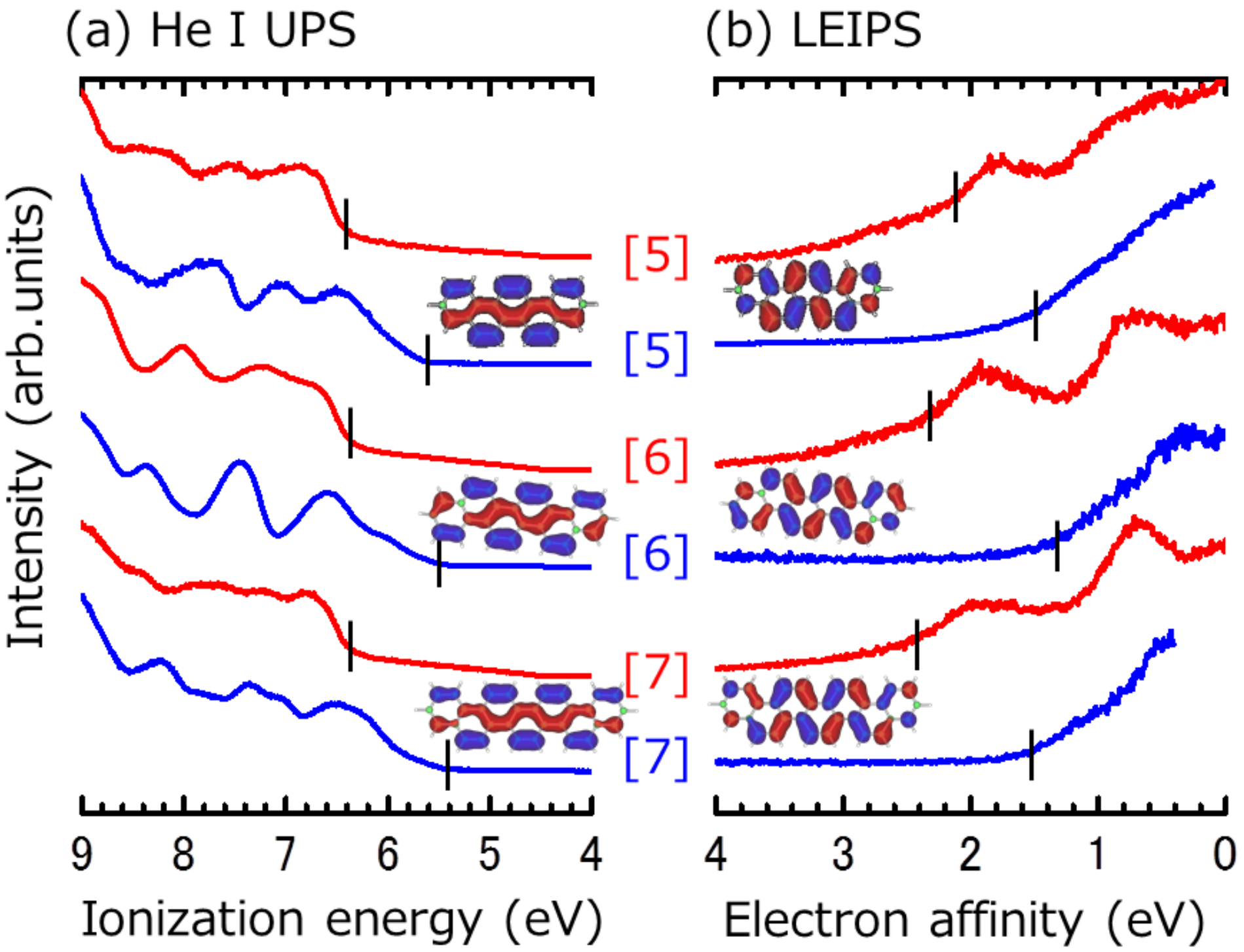

**Figure 3** UPS and LEIPS spectra of thin films on HOPG (red) and $SiO_2$ (blue) for picene ([5]-phenacene), [6]-phenacene, and [7]-phenacene.

**Table 1** Ionization energies, electron affinities, bandgaps, optical gaps, and exciton binding energies, of orientation-controlled phenacene thin films together with gas-phase ionization energies and electron affinities. Molecular orientations in the film, standing and lying, were controlled by the substrates of $SiO_2$ and HOPG, respectively.

| Molecule | Orientation / substrate | IE (eV) | EA (eV) | band gap (eV) | optical gap (eV) [37] | exciton binding energy (eV) | $IE_{gas}$ (eV) | $EA_{gas}$ (eV) |
|---|---|---|---|---|---|---|---|---|
| **Picene** | standing / $SiO_2$ | 5.67 | 1.52 | 4.15 | 3.3 | 0.9 | 7.55 | 0.5420± 0.0080 [65] |
| | lying / HOPG | 6.43 | 2.16 | 4.27 | | 1.0 | | |
| **[6]-phenacene** | standing / $SiO_2$ | 5.49 | 1.36 | 4.13 | 3.1 | 1.0 | 7.35 | 0.57[(a)] |
| | lying / HOPG | 6.42 | 2.30 | 4.12 | | 1.0 | | |
| **[7]-phenacene** | standing / $SiO_2$ | 5.44 | 1.53 | 3.91 | 3.1 | 0.8 | 7.23 | 0.635[(a)] |
| | lying / HOPG | 6.41 | 2.42 | 3.99 | | 0.9 | | |

(a) calculated by B3LYP/6-31+G(d)

**2.3 Gas-phase UPS**

Gas-phase UPS provides the isolated-molecule reference IE required to evaluate solid-state stabilization, namely the polarization energy. As shown in Figure 4, the gas-phase IE decreases monotonically with extension of the phenacene backbone: 7.55 eV for picene, 7.35 eV for [6]-phenacene, and 7.23 eV for [7]-phenacene. The measured IE of picene is in good agreement with the previously evaluated value of 7.51 ± 0.02 eV, which is based on multiple reported gas-phase measurements [66]. The slight decrease in IE with increasing molecular size parallels the trend observed for films with the same molecular orientation, reflecting the upward shift of the HOMO-derived level as the π-conjugated framework is extended. These gas-phase values establish the intrinsic chemical trend of the isolated molecules, against which the film-level shifts induced by molecular orientation and solid-state polarization can be evaluated.

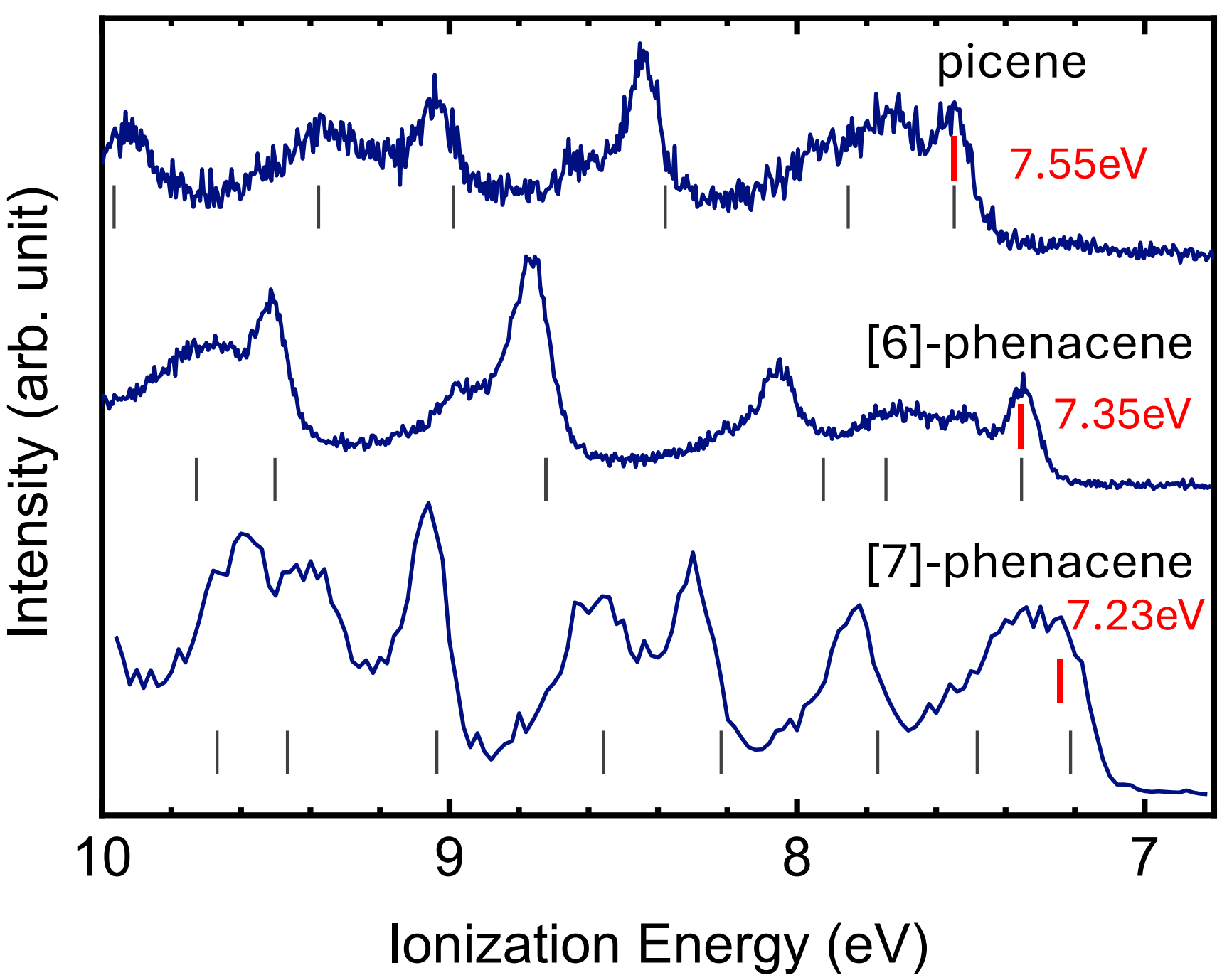

**Figure 4** Gas-phase UPS spectra of picene, [6]-phenacene, and [7]-phenacene. The black bars indicate the molecular orbital energies calculated using the DFT method. The vertical bars in red denote the HOMO peak positions of the spectra whose energies correspond to the gas-phase ionization energies (numbers in red).

**2.4 Polarization energy**

Charge carriers (holes and electrons) in organic semiconductors are mostly localized on a single or a few molecules and stabilized by the surrounding dielectric environment provided by the surrounding neutral molecules. This stabilization energy is referred to as the polarization energy. To clarify charge-carrier stabilization in the phenacene systems, its dependence on molecular length, and its relationship to the orientation-dependent energy levels, we evaluated the polarization energies for positive and negative charges, $P_+$ and $P-$, respectively, from the gas- and solid-state energy levels[67].

In this study, we include correction terms $\Delta^{\pm}$ that account for quantum-mechanical intermolecular interactions [35], and calculate the polarization energy, $P_{\pm}$, from Eq. (1)

$$P_+ = \mathrm{IE_{gas}} - \mathrm{IE} - \Delta^+,$$
$$P_- = -\mathrm{EA_{gas}} + \mathrm{EA} - \Delta^- \quad (1)$$

where $\mathrm{IE_{gas}}$ and IE are the gas- and solid-state ionization energies, respectively, and $\mathrm{EA_{gas}}$ and EA are the corresponding electron affinities. The quantum-mechanical correction terms were evaluated as one half of the total bandwidths. For this purpose, we used the energy-band structures reported by Nguyen *et al.* [68]. They reported band structures calculated using two types of crystal structures: one based on lattice constants fixed to the experimentally reported values (RPG) and the other based on fully optimized crystal structures including van der Waals interactions (vdW). Because the calculated band structure depends sensitively on the crystal structure, we carefully selected the appropriate bandwidths by examining the structural parameters used in the calculations.

For picene, Hosokai et al. reported that the UPS spectra of picene/$SiO_2$ are similar to those of the single crystal [58]. The single-crystal X-ray structures have been reported [69, 70]. Since the structural parameters used for the RPG band calculation are consistent with these single-crystal data, we adopted the RPG bandwidth for picene. For the higher phenacenes, single-crystal X-ray structural analysis appears to be difficult [71]. However, reliable lattice constants for thin films of [6]-phenacene and [7]-phenacene have been reported from X-ray diffraction measurements [59]. For [7]-phenacene, the reported structure is consistent with the RPG structure used by Nguyen *et al.*, and therefore we adopted the RPG bandwidth.

For [6]-phenacene, however, the RPG bandwidth calculated from the lattice constants obtained by powder X-ray diffraction [43] appears to be unrealistically narrow, and the corresponding cell volume is too large. In contrast, the fully optimized vdW structure for [6]-phenacene is close to the thin-film structure reported by Zwadlo *et al*[59]. Therefore, for [6]-phenacene, we adopted the bandwidth calculated for the vdW structure. The evaluated bandwidths, which correspond to twice the bandwidth correction terms $\Delta_{\pm}$, are summarized in Table 2.

Using these correction terms, we derived the polarization energies $P_+$ and $P_-$, as summarized in Table 2. The standing films on $SiO_2$ exhibit large hole polarization energies, $P_+$, of 1.64, 1.66, and 1.49 eV for picene, [6]-phenacene, and [7]-phenacene, respectively, whereas the lying films on HOPG show smaller values of 0.88, 0.73, and 0.52 eV, respectively. In contrast, the electron polarization energies, $P_-$, are 0.84, 0.68, and 0.74 eV on $SiO_2$, and 1.48, 1.62, and 1.63 eV on HOPG, for picene, [6]-phenacene, and [7]-phenacene, respectively. Thus, unlike $P_+$, $P_-$ is larger for the lying films on HOPG than for the standing films on $SiO_2$.

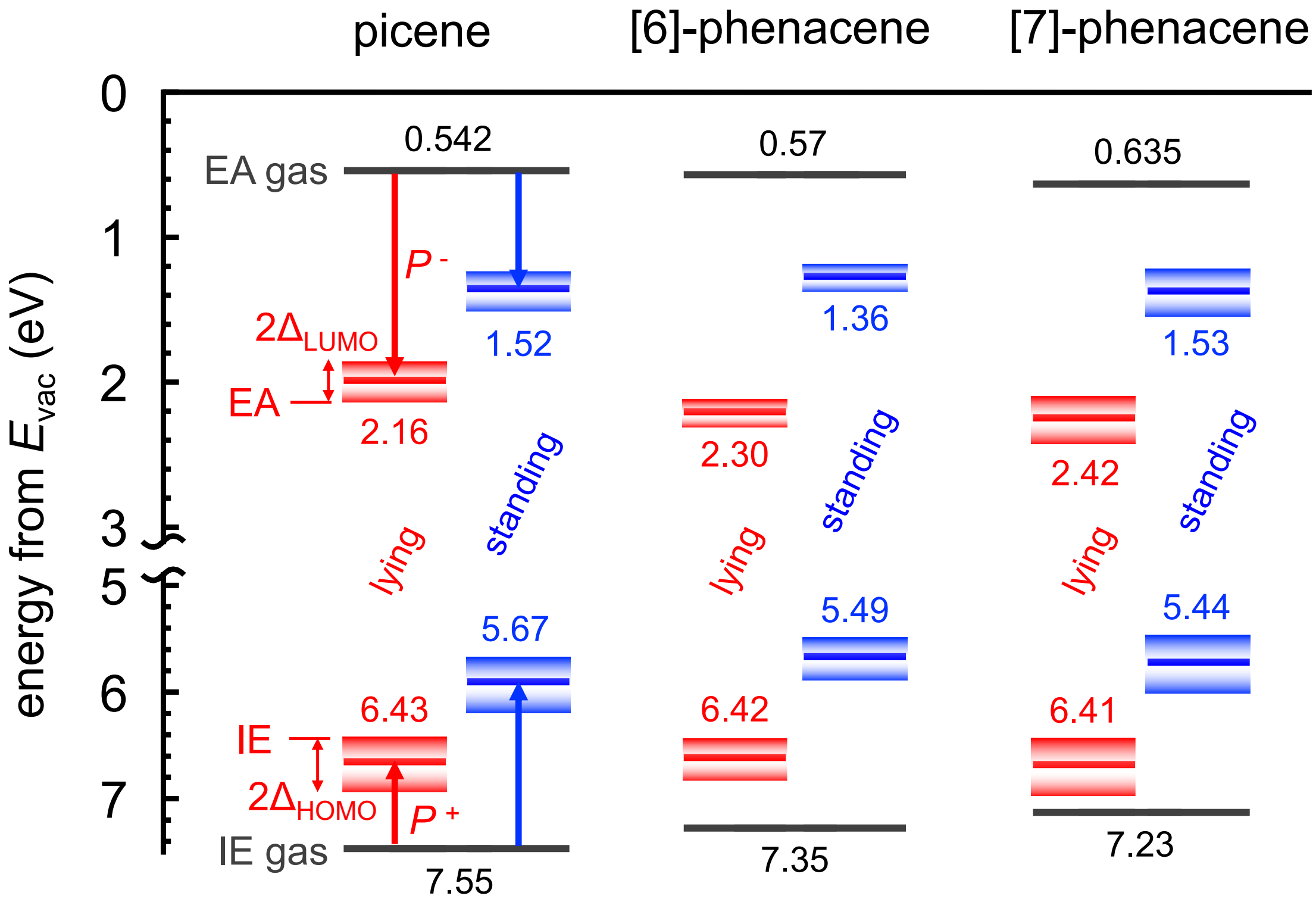

**Figure 5** Energy level diagram and polarization energies of picene, [6]-phenacene, and [7]-phenacene thin films. Ionization energies, IE, and electron affinities, EA, in the solid phase are compared with the corresponding gas-phase values, $IE_{gas}$ and $EA_{gas}$. Red and blue levels denote lying films on HOPG and standing films on SiO2, respectively. The polarization energies $P_{\pm}$ are evaluated from the differences between the gas- and solid-state energy levels, including quantum-mechanical correction terms. These correction terms are approximated as one half of the total bandwidths, $\Delta_{HOMO}$ and $\Delta_{LUMO}$, respectively.

**Table 2** Polarization energies $P_{\pm}$ and related energy parameters derived from IE(UPS)/EA(LEIPS) values and gas-phase IE/EA values.

| Molecule | Orientation/ substrate | polarization energy | | induction term | electrostatic term | bandwidth 2*Δ* | |
|---|---|---|---|---|---|---|---|
| | | $P_+$ (eV) | $P_-$ (eV) | *D* (eV) | *W* (eV) | HOMO (eV) | LUMO (eV) |
| **picene** | standing/ $SiO_2$ | 1.64 | 0.84 | 1.24 | 0.40 | 0.482 | 0.272 |
| | lying/ HOPG | 0.88 | 1.48 | 1.18 | -0.30 | | |
| **[6]-phenacene** | standing/ $SiO_2$ | 1.66 | 0.68 | 1.17 | 0.49 | 0.397 | 0.220 |
| | lying/ HOPG | 0.73 | 1.62 | 1.18 | -0.44 | | |
| **[7]-phenacene** | standing/ $SiO_2$ | 1.49 | 0.74 | 1.12 | 0.38 | 0.596 | 0.306 |
| | lying/ HOPG | 0.52 | 1.63 | 1.08 | -0.56 | | |

### 2.5 Origin of the orientation effect

To understand how the polarization energies depend on molecular orientation and charge-carrier polarity, we decomposed the polarization energy into two contributions [34, 72, 73]. The first contribution is the electronic polarization energy, or induction term, which stabilizes the charge carrier through the dielectric response of the surrounding medium. The second contribution is the electrostatic energy, or electrostatic term, which arises from the interaction between the charge carrier and the permanent charge distribution of the surrounding molecules. The polarization energy can therefore be expressed as the sum of the induction term, $D^{\pm}$, and the electrostatic term, $W^{\pm}$

$$P_{\pm} = D^{\pm} + W^{\pm} \tag{2}$$

Here, the plus and minus signs denote positive and negative charge carriers, respectively. Because the induction term originates from the induced response of the surrounding medium, its magnitude is assumed to be independent of the charge polarity, $D \equiv D^{+} = D^{-}$.

In contrast, $W$ represents the Coulomb interaction between the excess charge and the permanent charge distribution of the surrounding molecules; therefore, its sign depends on the charge polarity: $\mathrm{W} \equiv \mathrm{W}^{+} = -\mathrm{W}^{-}$. Combining these relations gives

$$\begin{aligned} P_{+} &= D + W \\ P_{-} &= D - W \end{aligned} \tag{3}$$

This formulation enables us to evaluate the induction and electrostatic contributions separately from the experimentally determined polarization energies, $P_{+}$ and $P_{-}$.

The evaluated $D$ and $W$ values are shown in Table 2. The induction term is approximately 1.1 eV and is almost independent of molecular orientation. In contrast, the electrostatic term depends strongly on molecular orientation: it is positive, in the range of approximately +0.4 to +0.5 eV, for the standing orientation, whereas it is negative, in the range of approximately −0.3 to −0.6 eV, for the lying orientation. The molecular-size dependences of both the induction and electrostatic terms are small. The large orientation dependence of the electrostatic term, together with the weak orientation dependence of the induction term, is consistent with previous results for pentacene [35], pentacenequinone [53], and perylene diamide [62].

The orientation-dependent electrostatic term originates from the interaction between the charge carrier and the permanent molecular quadrupole moment. We therefore calculated the quadrupole tensors using the B3LYP/6-31+G(d) method. The tensor components along the principal molecular axes are summarized in Table 3. As is commonly observed for unsubstituted polycyclic aromatic hydrocarbons, the largest tensor component is along the direction normal to the molecular plane. This is because the quadrupole moment originates primarily from the extended π-electron distribution of the molecular skeleton [74]. In this case, the magnitude of the quadrupole moment can be represented by the largest tensor component [75].

As shown in Figure 6(a), the quadrupole moment increases almost linearly with molecular size, namely with the number of fused benzene rings. Using the molecular volume evaluated based on the crystal structures [59], we also compared the quadrupole moment normalized by the molecular volume. The resulting value is nearly independent of molecular size (Figure 6(b)). This indicates that the quadrupole moment per unit molecular volume is nearly constant across the phenacene series.

Picene and [7]-phenacene possess permanent dipole moments along the molecular short axes, whereas [6]-phenacene does not. Because these phenacenes crystallize in the non-centrosymmetric space group $P2_1$ [43 76], the molecular dipole moments of picene and [7]-phenacene do not cancel in the crystal. Nevertheless, no anomaly attributable to the permanent dipole is observed for picene or [7]-phenacene in comparison with [6]-phenacene. This is most likely because the dipole moments of picene and [7]-phenacene are negligibly small, 0.0328 and 0.0361 D, respectively. Thus, the electrostatic term is predominantly governed by the permanent quadrupole moment. These results indicate that the quadrupole-derived electrostatic term is a key factor controlling the orientation-dependent frontier energy levels in phenacene thin films.

It is worth noting that predicting the electrostatic term W, as well as the gas-phase IE and EA, is essential for predicting solid-state frontier energy levels, namely the ionization energy and electron affinity in thin films, because these levels depend strongly on molecular orientation. In contrast to the electrostatic potential generated by a permanent dipole, which can be calculated based on the Helmholtz equation, W generated by the quadrupole moment is usually difficult to predict. For the lying orientation, however, we found that $W$ shows an approximately linear dependence on the molecular quadrupole moment, as shown in Figure 6(c). In contrast, no clear correlation is observed for the standing orientation, probably because the molecular tilt angle in the film varies with molecular size. These results suggest that orientation-dependent energy levels could be predicted in the future if the detailed thin-film structure can be predicted.

**Table 3** Quadrupole tensors calculated using the DFT method (B3LYP/6-31+G(d)).

| Molecule | Long-axis direction (Debye nm) | Short-axis direction (Debye nm) | Out-of-plane direction (Debye nm) |
|---|---|---|---|
| picene | 0.945 | 0.839 | -1.784 |
| [6]-phenacene | 1.113 | 0.995 | -2.109 |
| [7]-phenacene | 1.341 | 1.096 | -2.437 |

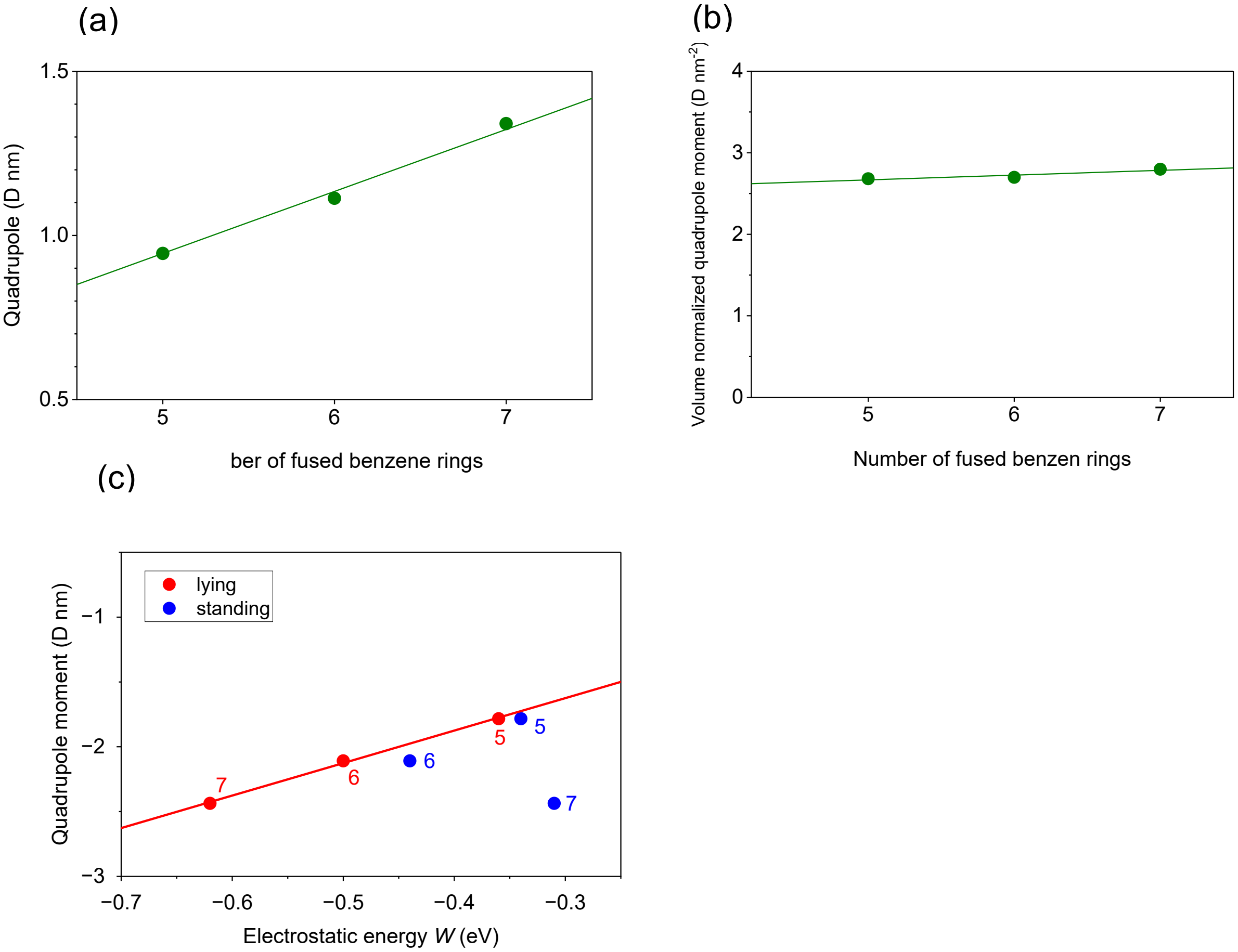

**Figure 6.** Molecular-size dependence of the quadrupole moment of phenacenes. (a) Quadrupole tensor component normal to the molecular plane as a function of the number of fused benzene rings. (b) Quadrupole moment normalized by the molecular volume. The quadrupole moment increases with molecular size, whereas the volume-normalized value is nearly constant. (c) Relationship between the calculated electrostatic energy and the experimentally determined electrostatic energy for the lying (red) and standing (blue) orientations. The numbers represent fused benzene rings. The solid line represents the best-fit linear relationship for the lying orientation with a slope of -2.5 $\pm$ 0.1 and an intercept of -0.87 $\pm$ 0.06.

## 3. Conclusion

We determined the electronic parameters, including the ionization energies, electron affinities, bandgaps, and exciton binding energies, of orientation-controlled thin films of picene, [6]-phenacene, and [7]-phenacene. The combination of MAES and UPS confirms lying orientations on HOPG and standing orientations on $SiO_2$, while UPS and LEIPS provide precise determinations of the ionization energies and electron affinities. The ionization energies and electron affinities depend only weakly on the number of fused benzene rings; their variation is less than 0.3 eV and closely follows the trend observed for the gas-phase ionization energies. In contrast, the absolute energy levels vary strongly, by approximately 1 eV, depending on molecular orientation.

Because the two frontier transport levels, the HOMO level (ionization energy) and LUMO level (electron affinity), shift in the same direction with changes in molecular size and orientation, the bandgaps remain nearly constant at approximately 4 eV. Similarly, the exciton binding energies are approximately 1 eV for all three phenacenes. As predicted in our previous study [50], the exciton binding energies are approximately one quarter of the corresponding bandgaps and one third of the optical gaps.

To understand the large orientation dependence of the frontier energy levels, we extended the analysis to the electron polarization energy. Using the gas-phase ionization energies and electron affinities, we found that $P_{-}$ is larger for the lying films on HOPG than for the standing films on $SiO_2$. This behavior contrasts with that of $P_{+}$ and demonstrates that cationic and anionic polarization energies must be evaluated independently. From the polarization energies $P_{+}$ and $P_{-}$, we derived the induction term (or electronic polarization term), $D$, and the electrostatic term, $W$, following the procedure reported previously [35]. The induction term is approximately 1 eV and increases slightly with molecular size, but it shows little dependence on molecular orientation. In contrast, the electrostatic term $W$ depends strongly on molecular orientation. We further confirmed that $W$ originates primarily from the permanent molecular quadrupole moment normal to the molecular plane rather than the permanent dipole moment.

These results position phenacenes as experimentally accessible finite benzenoid models of graphene-nanoribbon substructures and demonstrate that molecular orientation provides a powerful means of tuning charged-state stabilization without substantially altering the bandgap.

## 4. Methods

Picene, [6]-phenacene, and [7]-phenacene were prepared by NARD Institute, LTD. Thin films were fabricated by vacuum deposition onto HOPG and $SiO_2$ substrates under ultra-high-vacuum conditions. HOPG substrates were cleaved in air and cleaned by heating at 673 K in ultra-high vacuum. SiO2 substrates, prepared as native oxide layers on Si(111), were cleaned by ultrasonication in acetone, isopropanol, and ultrapure water, followed by heating at 673 K in ultra-high vacuum. Surface cleanliness was verified by UPS and MAES before film deposition. The phenacene molecules were deposited at a rate of 0.1–0.2 nm $min^{-1}$ at pressures below $1.1 \times 10^{-7}$ Pa. The average film thicknesses were 5–20 nm for the standing orientation and 1–2 nm for the lying orientation, corresponding to approximately 5–10 monolayers in both cases. No discernible thickness dependence was observed in the spectra, confirming that the obtained IE and EA values are representative of bulk-like films.

Gas-phase UPS was measured to determine the isolated-molecule ionization energies using a VG CLAM4 electron analyzer with He I light. The sample molecules were sublimed from a Mo oven mounted on a light source head. The gas-phase data were calibrated using the Ar $2p_{3/2}$ line at 15.76 eV, which was measured simultaneously. Regarding gas-phase electron affinities, no experimental gas-phase EAs were available for [6]-phenacene and [7]-phenacene. Thus, the EAs were calculated with Gaussian 16 at the B3LYP/6-31+G(d) level. The adiabatic electron affinity EA was obtained from the total-energy difference between the optimized neutral molecule and the optimized radical anion. Frequency calculations were used to confirm that the optimized structures corresponded to minima.

Thin-film He I UPS and MAES were performed in situ. UPS spectra were measured using a PHOIBOS-HSA100 hemispherical analyzer with an energy resolution of approximately 60 meV and an electron acceptance angle of ±9 degrees. He I UPS was recorded with a photon incidence angle of 45 degrees and an electron emission angle of 0 degrees. He*($2^3S$) MAES was recorded with a metastable helium beam incidence angle of 0 degrees and electron an electron emission angle of 60 degrees. The vacuum level in UPS was obtained from the secondary-electron cutoff under a -5.0 V sample bias. Details of the LEIPS apparatus have been reported elsewhere [77]. The LEIPS measurements were also performed in situ after film preparation under ultra-high vacuum. For photon detection, a bandpass filter with a center wavelength of 285 nm (BrightLine, Semrock) was used in combination with a photomultiplier tube (R585s, Hamamatsu, K.K.). The LEIPS energy resolution was approximately 300 meV, estimated from the Fermi edge of a silver thin film. The vacuum level was derived from the first derivative of low-energy electron transmission (LEET) spectra. All thin-film measurements were performed at room temperature.

**Acknowledgements**
The authors thank Professor Yasushi Nishihara of Okayama University for providing the materials used in this study. The authors also gratefully acknowledge Mr. Yousuke Suda and Mr. Takuma Yamaguchi for their valuable assistance with the measurements. This work was partially supported by JSPS KAKENHI Grant Numbers JP24H00446, JP26248062, and JP23H05461. M.K. acknowledges support from the JSPS Research Fellowship for Young Scientists (Grant Number JP23KJ0310).

**Author contributions**
S.K. conceived the project. R.M., K.Y., and S.K. performed the UPS, MAES, and LEIPS measurements and analyses, with assistance from H.Y. in the LEIPS measurements. M.K. performed DFT calculations. M.K. and H.Y. analyzed the polarization energies. R.M. prepared the initial draft of the manuscript under the supervision of S.K. H.Y. wrote and finalized the manuscript. All authors discussed the results and approved the final version of the manuscript.

**Data Availability Statement**

The data that support the findings of this article are available from the corresponding author upon reasonable request.

**Conflicts of Interest**

The authors have no conflicts of interest to disclose.